\documentstyle[preprint,aps]{revtex}

\begin{document}

\title{Gravitational lensing statistical properties
                in general FRW cosmologies with dark energy component(s):
                analytic results}

\author{Zong-Hong Zhu\thanks{e-mail address: zhuzh@bao.ac.cn} 
	}
\address{Beijing Astronomical Observatory,
                Chinese Academy of Sciences, Beijing 100012, China\\
        National Astronomical Observatories,
                Chinese Academy of Sciences, Beijing 100012, China
        }

\maketitle

\begin{abstract}

Various astronomical observations have been consistently making a strong
case for the existence of a component of dark energy with
negative pressure in the universe.
It is now necessary to take the dark energy component(s) into account
in gravitational lensing statistics and other cosmological tests.
By using the comoving distance we derive analytic but simple
expressions for the optical depth of multiple image,
                 the expected value of image separation and
                the probability distribution of image separation
caused by an assemble of singular isothermal spheres
in general FRW cosmological models with dark energy component(s).
We also present the kinematical and dynamical properties of these
kinds of cosmological models and calculate the age of the universe
and the distance measures, which are often used in classical cosmological
tests.
In some cases we are able to give formulae that are simpler than those
found elsewhere in the literature, which could make the cosmological tests for
dark energy component(s) more convenient.

\end{abstract}

\newpage

\section{Introduction}

The Hubble expansion, the Cosmic Microwave Background Radiation(CMBR)
and the primordial Big Bang Nucleosynthesis(BBN) are three cornerstones
of the standard hot Big Bang cosmological model.
Recently these three kinds of observations have been making a strong
case for the existence of a nearly uniform component of dark energy with
negative pressure:
        More than twenty experiments of CMBR measurements have now covered three
        orders of magnitude in multipole number, and are beginning to
        define the position of the first acoustic peak at a value that is
        consistent with a flat universe\cite{lin98};
            However, the deuterium abundance measured in four high redshift
            hydrogen clouds seen in absorption against distant quasars
            \cite{bur98} combined with baryon fraction in galaxy
            clusters from X-ray data\cite{whi93}
            give a low matter density universe,
            $\Omega_m \sim 0.3$---$0.4$;
        The discrepancy between the low matter density and a flat universe
        is observationally resolved by high redshift Ia type supernovae (SNeIa)
        observations\cite{per98rie98}, which
        implies an accelerating universe driven by a dark, exotic form of
        energy component.
    For a recent review of observational evidence for the dark
    energy component, reader is referred to Ref.~\cite{tur98}.
It seems that determining the amount and nature of the dark energy
is emerging as one of the most important challenges in cosmology.

The simplest possibility for the dark energy component is cosmological
constant $\Lambda$ or equivalently vacuum energy\cite{wei89}.
In the past, a  nonzero cosmological constant has been advocated and then
with the improved observational data discarded several times\cite{coh98sah99}.
Due to this checked history and the difficulty in understanding the
observed $\Lambda$ in the framework of modern  quantum field theory,
now most physicists and astronomers prefer other candidates for dark energy,
including a frustrated network of
topological defects such as strings or walls\cite{defect}
and an evolving scalar field referred to
by some as quintessence\cite{quint} etc.
As shown in literatures, it is difficult to discriminate against
these different possibilities either by the SNeIa data alone\cite{gar98}
or only by the CMBR data\cite{hue99}. This led some authors
to consider the combination of the SNeIa measurements with the anisotropy
of CMBR\cite{tur97} or the large scale structures\cite{per99}.

Because of the potential importance of probing the dark energy component
and uncertainties of present constraints, it is worthy of examining various
cosmological tests for dark energy component(s).
It has been long shown that gravitational lensing statistics is an
efficient tool for determining cosmological parameters\cite{lensing,arcs}.
Some authors have even given the general expressions for the optical depth
and average image separation in general Friedman-Robertson-Walker (FRW)
cosmological models\cite{analytic}.
But these expressions are complicated and thereby hard to apply in practice.
Furthermore, they didn't involve general dark energy component(s).
We reconsider here gravitational lensing statistical properties,
such as the optical depth of multiple image,
                 the expectation value of image separation and
                the probability distribution of image separation,
in general FRW cosmologies with various matter and energy components.
We also calculate the age of the universe and distance measures, such as
the luminosity distance and the angular diameter distance, which are often
involved in classical cosmological tests.
In some cases we are able to give formulae that are simpler than those
found elsewhere in the literature, which could make the cosmological tests for
dark energy component(s) more convenient.

The paper is organised as follows:
In section 2, we summarize kinematics and dynamics of general FRW cosmologies
with dark energy component(s);
In section 3, we derive expressions for the age of the universe and the
various distance measures;
In section 4, using the comoving distance, we generalize and simplify
lensing formulae;
Results and discussion are given in section 5.

\section{Kinematics and dynamics of general FRW cosmologies
                with dark energy component(s)}

For general FRW cosmologies, the metric of
spacetime is described by (in the $c = 1$ unit):
\begin{equation}
ds^2 = -dt^2 + R^2(t) \left[ d\chi^2 +
f^2(\chi)(d \theta^2 + \sin^2\theta d\phi^2)\right] \,\,,
\label{metric}
\end{equation}
where $f(\chi) = \chi$ for a flat universe ($k=0$), $f(\chi) = \sin\chi$
for a closed universe ($k= + 1$), and $f(\chi) = \sinh\chi$ for an open
universe ($k= - 1$).
Defining the scale factor $a(t) = R(t)/R_0$,  $a=1$ today, and
the Friedmann equation takes the form
\begin{equation}
\begin{array}{cl}
\left(\frac{\dot{a}}{a} \right)^2 & =
\frac{8 \pi G}{3}\sum_i \rho_i - \frac{k}{a^2 R_0^2}  \\
 & \\
\frac{\ddot{a}}{a}   & =
- \frac{4 \pi G}{3}\sum_i (\rho_i + 3p_i)
\end{array}
\label{friedmann}
\end{equation}
where the dot represents derivatives with respect to $t$, and
$i$ includes all components of matter or energy in the universe,
e.g., $i=m$ for the total nonrelativistic matter,
        $r$ for the total radiation component,
  $\Lambda$ for the cosmological constant
                                ($\rho_\Lambda \equiv \Lambda / (8\pi G)$),
        $c$ for the cold dark matter,
        $h$ for the hot dark matter,
        $b$ for the baryons and
        $x$ for some unknown exotic component.
If the effective equation of state for the $i$-th component is parameterized
as $\omega_i = p_i/\rho_i$, its density scales as $\rho_i \propto a^{-n_i}$
where $n_i = 3(1+\omega_i)$.
For instance, nonrelativistic matter scales
as $\rho_m \propto a^{-3}$ while relativistic matter, such as radiation,
changes as $\rho_r \propto a^{-4}$, and vacuum energy(cosmological constant)
is invariant ($\rho_{\Lambda} \propto a^0$) as the universe expands.
Eq.(\ref{friedmann}) tells us that all components of matter or energy and
a curvature term conspire to drive the universal expansion.
It is convenient to assign symbols to their respective fractional
contributions at the present epoch.
Defining parameters
\begin{equation}
\Omega_i =  \frac{8 \pi G }{3 H_0^2} \rho_{i0}, \,\,\,\,\,\,
\Omega_k =  \frac{-k}{R_0^2 H_0^2}  \,\,,
\label{omega}
\end{equation}
where $H_0$ is the Hubble constant and zero subscripts refer to the
present epoch.
Eq.(\ref{friedmann}) implies a relation
\begin{equation}
1 = \sum_i \Omega_i + \Omega_k  \,\,.
\end{equation}
The parameters $\Omega_i, \Omega_k$ can be used to rewrite equation
(\ref{friedmann}) at a general time
\begin{equation}
\frac{1}{a }\frac{da}{dt}  = H_0 \left(\sum_i \Omega_i a^{-3(1+\omega_i)}
                  +\Omega_k a^{-2} \right)^{1/2} \,\,.
\label{dadt}
\end{equation}

\section{The age of the universe and distance measures}

Equation (\ref{dadt}) can be easily used to calculate the age of the
universe. The age is the integral of the time up to the present epoch 
in terms of the normalized scale factor $a$:
\begin{equation}
H_0 t_0 = \int_0^1 \frac{da}{a\left[\sum_i \Omega_i a^{-3(1+\omega_i)}
                                + \Omega_k a^{-2}\right]^{1/2}}  \,\,.
\label{H0t0a}
\end{equation}
By a trivial change of the integral variable from $a$ to redshift $z$,
the age expression reduces to
\begin{equation}
H_0 t_0 = \int_0^{\infty} \frac{dz}{(1+z) \left[ \sum_i \Omega_i
                (1+z)^{3(1+\omega_i)} + \Omega_k (1+z)^2 \right]^{1/2}}  \,\,.
\label{H0t0z}
\end{equation}

It has been shown that the natural cosmological distance for the analysis
of gravitational lensing statistics is the comoving distance\cite{zhu98}.
The distance a light ray travels can be calculated as follows.
Light rays follow null geodesics where
$ds^2 =0$ so that $dt^2 = R_0^2 a^2 d\chi^2$.
With Eq.(\ref{friedmann}) and the relation $a(t) = (1+z)^{-1}$, the comoving
distance is
\begin{equation}
\chi = \left\{ \begin{array}{ll}
                \int_0^z \frac{dz}{\sqrt{\sum_i \Omega_i (1+z)^{3(1+\omega_i)}}}
                                &       (k=0)\,,\\
                & \\
                \left| \Omega_k \right|^{1/2} \int_0^{z}
                         \frac{dz}{\sqrt{\sum_{i} \Omega_i (1+z)^{3(1+\omega_i)}
                        + \Omega_k (1+z)^2}}           & (k=\pm 1)\,.\\
                \end{array}
        \right.
\label{comoving}
\end{equation}
Most of classical cosmological tests involve the standard candle (e.g.
SNeIa) or standard ruler (e.g. radio source). Therefore both luminosity
distance $D^L$ and angular diameter distance $D^A$ are often used.
They are related to the comoving distance as\cite{CPT}
\begin{equation}
D^L \equiv R_0^2 f(\chi) / R(t) ,\,\,\,\,\,\,
D^A \equiv R(t) f(\chi).
\end{equation}
They are simply related to the measurable, the redshift $z$, through
Eq.(\ref{comoving}) with the following form
\begin{equation}
D^L = (1+z)^2 D^A
    =   \left\{ \begin{array}{ll}
                H_0^{-1} (1+z) \chi & k=0,\\
                H_0^{-1} (1+z) \left|\Omega_k\right|^{-1/2} \sin \chi
                                                \,\,\,\,\,      & k=+1,\\
                H_0^{-1} (1+z) \left|\Omega_k\right|^{-1/2} \sinh \chi
                                                                & k=-1,\\
              \end{array}
        \right.
\label{dLdA}
\end{equation}
Finally, one often needs to calculate volume element of the universe.
In order to take into account the cosmic expansion, it is more convenient
to use comoving volume than the traditional physical volume.
Within the shell $d\chi$ at $\chi$, $dV$ reads
\begin{equation}
dV = 4\pi R^3_0 f^2(\chi)d\chi = 4\pi H_0^{-3}\left\{ \begin{array}{ll}
                        \chi^2 d\chi                   & k=0,\\
                        \left|\Omega_k\right|^{-3/2} \sin^2 \chi d\chi & k=+1,\\
                        \left|\Omega_k\right|^{-3/2} \sinh^2\chi d\chi & k=-1.\\
                \end{array}
        \right.
\end{equation}

\section{Gravitational lensing statistical properties of singular
                isothermal spheres}

        \subsection{Optical depth for multiple image}

First of all, we consider the lensing cross-section\cite{tog84}
 due to a specific galaxy.
Following Ref.~\cite{tog84}, we model the mass density profile of the total
galaxy matter as the singular isothermal sphere (SIS).
The dimensionless cross-section of multiple image for a point source located
at $z_s$ produced by a single SIS galaxy at $z_d$ is\cite{SEFWu}
\begin{equation}
        \hat{\sigma} = \pi \theta^2_E, \,\,\,\,
                        \theta_E \equiv 4\pi \sigma^2
                                        \frac{D^A_{ds}}{D^A_s} \,,
\end{equation}
where $D_s^A$ and $D_{ds}^A$ are the angular diameter distances from the
observer to the source and from the lens to the source respectively,
$\theta_E$ is the angular radius of Einstein ring and
$\sigma$ is the velocity dispersion of the lensing galaxy.
Using the comoving distance gives
\begin{equation}
        \hat{\sigma} = 16{\pi}^3 \sigma^4
                        \left[\frac{f(\chi_s - \chi_d)}{f(\chi_s)}\right]^2
\,.
\end{equation}

Now, let's consider the contributions of an ensemble of
galaxies having different luminosities and redshifts.
The present-day galaxy luminosity function can be described by the Schechter
function\cite{pee93}
\begin{equation}
\phi_i(L)dL=\phi_i^*(L/L_i^*)^{-\alpha_i}\exp(-L/L_i^*)d(L/L_i^*),
\label{schechter}
\end{equation}
where $i$ indicates the morphological type of galaxies: $i$=(E, S0, S).
The above expression can be converted into the velocity dispersion distribution
through the empirical formula between the luminosity and the central dispersion
of local galaxies $L/L_i^*=(\sigma/\sigma_i^*)^{g_i}$.
Finally, the optical depth of multiple image by galaxies at redshifts
ranging from 0 to $z_s$ for the distant sources like quasars at $z_s$ is
\begin{equation}
    \tau(z_s) = \left( \sum_{i = E, S0, S} F_i \right)    T(z_s)
\,,
\label{tau-frame}
\end{equation}
The parameter $F_i$ represents the effectiveness of the $i$-th morphological
type of galaxies in producing double images\cite{tog84}, which reads
\begin{equation}
F_i \equiv 16\pi^3 {H_0}^{-3}
                        \langle n_{0i} {\sigma}^4 \rangle
        = 16\pi^3 {H_0}^{-3} \phi_i^* \gamma_i
                \left( b_i \sigma_i^* \right)^4
                \int (L/L_*)^{\alpha_i + 4/g_i} \exp(L/L_*)dL/L_*
\,,
\label{Fi-frame}
\end{equation}
where $\gamma_i$ is the galaxy morphological composition and $b_i$ is the
velocity bias between the velocity dispersion of stars and of dark matter
particles\cite{koc96}.
The above equation can be further written as
\begin{equation}
F_i = 16\pi^3 {H_0}^{-3} \phi_i^* \gamma_i
        \left( b_i \sigma_i^* \right)^4 \Gamma(-\alpha_i+4/g_i+1), \,\,\,
        {\rm if} \,\,\, L \in (0,\infty)
\,,
\label{Fi-max}
\end{equation}
if the integral is performed from $0$ to $\infty$.
In practice, the galaxy luminosities have the minimum and maximum limits,
and, therefore, Eq.(\ref{Fi-max}) is the maximum estimate of $F_i$.
The $z_s$ dependent factor $T(z_s)$  is
\begin{equation}
        T(z_s) = \left( H_0 R_0 \right)^3
                \int_0^{\chi_s} {\left[ \frac{f(\chi_s - \chi_d)}
                {f(\chi_s)} \right]}^2 f^2(\chi_d) d\chi_d
\,.
\label{T-frame}
\end{equation}
For general FRW cosmologies with various matter and energy components,
an analytic expression is found\cite{zhu98}:
\begin{equation}
T(z_s) = \left\{ \begin{array}{ll}
         \frac{\chi^3_s}{30}, & (k=0)\,,\\
        \left|\Omega_k \right|^{-3/2}
        \left[\frac{1}{8}(1 + 3 \cot^2\chi_s) \chi_s -
        \frac{3}{8}\cot\chi_s\right], & (k=+1)\,,\\
        \left|\Omega_k \right|^{-3/2}
        \left[\frac{1}{8}(-1 + 3 \coth^2\chi_s) \chi_s -
       \frac{3}{8}\coth\chi_s\right], \,\,\, & (k=-1)\,,\\
        \end{array}
        \right.
\label{T-general}
\end{equation}
where $\chi_s$ can be calculated through Eq.(\ref{comoving}).

As pointed out by  Kochaneck\cite{koc96}, although the parameters
($\phi_i^*, L^*_i, \alpha_i; \sigma_i^*, g_i; \gamma_i$)
of galaxies are updated due to recent surveys\cite{lov92}
there are still large uncertainties with them.
Therefore we treat $F \equiv \sum_i F_i$ as normalized factor.
One may expect to determine both the dark energy amount and its equation
of state. It is necessary to consider general flat universe models
($\Omega_m; \Omega_x, \omega_x$) with $\Omega_m + \Omega_x =1$.
Generally, the optical depth of gravitational lensing depends on the amount
of dark energy $\Omega_x$ as well as its equation of state $\omega_x$.
The larger the dark energy amount is, the higher the gravitational lensing
probability will be; the more negative the dark energy pressure is, the higher
the optical depth will be.

        \subsection{Average image separation}

In addition to optical depth, other interesting quantities for lensing
statistics are the mean image separation produced by a specified SIS
at a given $z_d$ with velocity dispersion $\sigma$ and its average
over the distributions of lens velocity dispersion and lens redshift.
The angular separation of the two images produced by SIS is
\begin{equation}
\Delta \theta = 2\theta_E
              = 8 \pi \sigma^2 \frac{f(\chi_s - \chi_d)}{f(\chi_s)}
\end{equation}
which doesn't depend on impact parameter, and hence implies
\begin{equation}
\overline {\Delta \theta} = \Delta \theta\,\,\,\,\,\,\,\,\, {\rm for ~~SIS}
\end{equation}
where $\overline {\Delta \theta}$ is the mean image separation for
the SIS lens at $z_d$ with velocity dispersion $\sigma$.
The average image separation is obtained by
\begin{equation}
< \overline {\Delta \theta} > = \frac{1}{\tau}
                                \int \overline {\Delta \theta} \,\, d\tau
\end{equation}
We take into account the lens redshift distribution as well as the lens
velocity dispersion distribution. The latter was generally neglected in
literatures. Then the average image separation reduces to
\begin{equation}
< \overline {\Delta \theta} > = \frac{1}{\tau}
                 \left( \sum_{i = E, S0, S} F_i^{\prime} \right)  \Theta(z_s)
\,,
\end{equation}
where
\begin{equation}
F_i^{\prime} \equiv 128\pi^4 {H_0}^{-3}
                        \langle n_{0i} {\sigma}^6 \rangle
        = 128\pi^4 {H_0}^{-3} \phi_i^* \gamma_i
                \left( b_i \sigma_i^* \right)^6
                \int (L/L_*)^{\alpha_i + 6/g_i} \exp(L/L_*)dL/L_*
\,,
\label{Fiprime-frame}
\end{equation}
if the above integral is performed from $0$ to $\infty$, it can be further
written as
\begin{equation}
F_i^{\prime} \equiv 128 \pi^4 {H_0}^{-3} \phi_i^* \gamma_i
                \left( b_i \sigma_i^* \right)^6 \Gamma(-\alpha+6/g_i+1), \,\,\,
                {\rm if} \,\,\, L \in (0,\infty)
                \,,
\label{Fiprime-max}
\end{equation}
and the $z_s$ dependent factor $\Theta(z_s)$ is
\begin{equation}
\Theta(z_s) \equiv \left( H_0 R_0 \right)^3
                \int_0^{\chi_s} {\left[ \frac{f(\chi_s - \chi_d)}
                {f(\chi_s)} \right]}^3f^2(\chi_d) d\chi_d
\,.
\label{theta-frame}
\end{equation}
For general FRW cosmologies with various matter and energy components,
an analytic expression is found:
\begin{equation}
\Theta(z_s) = \left\{ \begin{array}{ll}
         \frac{\chi^3_s}{60}, & (k=0)\,,\\
        \left|\Omega_k \right|^{-3/2} \frac{2}{15} \sin^{-3}\chi_s
         \left[ (4-3\sin^2 \chi_s) + \cos\chi_s(-4+\sin^2\chi_s) \right],
                         & (k=+1)\,,\\
        \left|\Omega_k \right|^{-3/2} \frac{2}{15} \sinh^{-3}\chi_s
         \left[ (-4-3\sinh^2 \chi_s) + \cosh\chi_s(4+\sinh^2\chi_s) \right],
                         & (k=-1)\,,\\
        \end{array}
        \right.
\label{theta-general}
\end{equation}
where $\chi_s$ can be calculated through Eq.(\ref{comoving}) as before.
If we define $\Delta \theta_* \equiv \sum_i F_i^{\prime} /\sum_i F_i$,
which is determined completely by the intrinsic characteristic parameters of
galaxies describing its statistical properties, the average image separation
then becomes
\begin{equation}
< \overline {\Delta \theta} > = \Delta \theta_* \, \Theta(z_s) / T(z_s) ,
        \,\,\,\,\,\,\,\,
        \Delta \theta_* \equiv \left( \sum_{i=E, S0, S} F_i^{\prime} \right)
                \div \left( \sum_{i=E, S0, S} F_i \right) .
\label{delta-theta}
\end{equation}
Thus for an open or closed universe, the average image separation for the
SIS lens depends on $\chi_s$ (or the source redshift), while for a flat
universe, it doesn't but takes a constant value
$< \overline {\Delta \theta} > = \Delta \theta_* / 2$.
This fact has been suggested to test directly the curvature of the
universe\cite{par97}.

        \subsection{Probability distribution of image separation}

Another useful quantity of lensing statistics is the probability distribution
of image separation, which is reconsidered here for general cosmologies
with various matter and energy components.
The differential optical depth for multiple imaging a background object
like a quasar by lenses at position between $\chi_d$ and $\chi_d + d\chi_d$
and with luminosity from $L$ to $L + dL$ is
\begin{equation}
d^2\tau_i = \gamma_i \phi_i(L)dL \,\, \hat{\sigma} \,\,
                R^3_0 f^2(\chi_d)d\chi_d .
\end{equation}
By change of the variable from luminosity to angular separation, the
differential optical depth becomes
\begin{equation}
\begin{array}{ll}
\frac{d^2\tau_i}{ d \Delta \theta d\chi_d} = &
              \pi^2 (8\pi)^{g_i(\alpha_i -1)/2 -1} \gamma_i g_i \phi_{*i}
                (b_i \sigma_{*i})^{g_i(\alpha_i -1)}    
		S(\Delta \theta) \, \Delta \theta ^{g_i(1-\alpha_i)/2 +1}\\
 & \\
 &		\cdot
                \exp \left\{-[8\pi (b_i \sigma_{*i})^2]^{-g_i/2}
                    \left[\frac{f(\chi_s-\chi_d)}{f(\chi_s)}\right]^{-g_i/2}
                        \, \Delta \theta ^{g_i/2} \right\}
              \left[\frac{f(\chi_s-\chi_d)}{f(\chi_s)}\right]^{g_i(\alpha_i-1)}
                R_0^3 f^2(\chi_d)                           \\
\end{array}
\end{equation}
where we have taken into account the angular resolution bias by the function
$S(\Delta \theta)$, which varies for different lens observations and
surveys\cite{jau95}.
Then, the probability distribution of image separations becomes
\begin{equation}
  \frac{dP}{d\Delta\theta} = \frac{1}{\tau} \int_{0}^{\chi_s} \sum_{i}
                \left(\frac{d^2\tau_i}{d\Delta\theta d\chi_d}\right) d\chi_d
\label{P-theta}
\end{equation}
which in general case depends on $\chi_s$ (or $z_s$, the redshift of the
source). However for a flat universe, the above integral scales as
$\propto \chi_s^3$. Combining Eq.(\ref{tau-frame}) and \ref{T-general},
one concludes that the probability distribution of image separations
is independent of $\chi_s$, regardless of the functional form of the
angular resolution bias\cite{GPL}.
Hence the cumulative distribution of image separations can be easily
calculated from $dP/d\Delta\theta$ as
\begin{equation}
P(\Delta\theta) = \int_{0}^{\Delta\theta}
                      \left(\frac{dP}{d\Delta\theta'}\right) d\Delta\theta'
\end{equation}

\section{Results and Discussions}

As shown above, it is easy to analytically calculate the optical depth
for multiple image (Eq.(\ref{tau-frame}) and (\ref{T-general})), average
image separation (Eq.(\ref{delta-theta}),  (\ref{T-general}) and
(\ref{theta-general})) and probability distribution of image separations
(Eq.(\ref{P-theta})), as well as the age of the universe (Eq.(\ref{H0t0z}))
and the luminosity (angular diameter) distance (Eq.(\ref{dLdA})),
if the matter and energy components in the universe are given in terms of
($\Omega_i, \omega_i$).
It is shown that the lensing statistical properties depend sensitively
on the dark energy amount and its equation of state and hence provide
an independent probe for the dark energy.

In the above lensing calculations, we have assumed that the comoving number
density of galaxies is constant. However, this may not hold true for the
realistic situation. The influence of galaxy evolution on the
lensing statistics should also be taken into account\cite{mao94}.
We have considered this effect by using the galaxy merging
model proposed by Broadhurst et al. (1992), since
the scenario of galaxy merging can account for both
the redshift distribution and the number counts of galaxies at
optical and near-infrared wavelengths\cite{merging}.
There are two effects arising from the galaxy merging:
The first  is that there are more galaxies and hence more lenses in the past.
The second  is that galaxies are typically less massive in the past and
hence less efficient as lenses.
As a result of two effects, the total  cross-section remains roughly
unchanged\cite{zhu97}.

Although the evidence for the existence of a nearly uniform component of
dark energy with negative pressure in the universe has been consistently
provided by several astronomical observations, we at present know a little
about it.
It needs more cosmologcial constraints on dark energy to determine
its amount and nature.
The simple formulae we derived here could make the cosmological tests for
dark energy component(s) more convenient.

\begin{acknowledgements}
I thank X. P. Wu for careful reading of the manuscript and helpful discussions.
This work was supported by the National Natural Science Foundation of China,
under Grant No. 19903002.
\end{acknowledgements}

\end{document}